# On the Conservativeness of Robust Variance Estimators in Propensity Score Weighted Cox Models

Hiroya Morita[1]*, Shunichiro Orihara[1], Fumitaka Shimizu[2,3], Masataka Taguri[1]

[1]Department of Health Data Science, Tokyo Medical University, Tokyo, Japan

[2]Department of Urology, Juntendo University Graduate School of Medicine, Tokyo, Japan

[3]Department of Urology, Juntendo Shizuoka Hospital, Shizuoka, Japan

*Correspondence:

Hiroya Morita, Department of Health Data Science, Tokyo Medical University, Tokyo, Japan

E-mail: d125035@tokyo-med.ac.jp

Short running tittle: Conservativeness of Robust Variance in Propensity Score Cox Models

**Fundings** This work was partially supported by JSPS KAKENHI Grant Number JP24K14862.

**Abstract**

In propensity score weighted analysis, robust variance that does not account for weight estimation is commonly used. In propensity score weighted Cox models (CoxPSW), the robust variance is known to be conservative when weights for the average treatment effect (ATE) are used, but it remains unclear whether this conservativeness also holds for other weighting schemes. This study evaluated the performance of the robust variance in CoxPSW when weights other than ATE are applied. We conducted an asymptotic comparison between the robust variance and a variance estimator that accounts for weight estimation under non-ATE weights. Their performance was further evaluated through simulation studies and real data analysis. The analytical results, simulations, and real data analysis indicated that the robust variance is not necessarily conservative in CoxPSW when weights other than ATE are used. These findings suggest that variance estimators that account for weight estimation should be used when applying non-ATE weights in CoxPSW.

# 1. Introduction

Observational studies using real-world data have become increasingly common. Because confounding is a major concern in observational research, it is essential to apply appropriate methods to address it. Propensity score weighted (PSW) analysis is one of the most widely used approaches [1,2].

PSW typically employs weights derived from propensity scores. By applying different weights, PSW enables the estimation of various causal estimands, including the Average Treatment Effect (ATE), the Average Treatment Effect on the Treated (ATT), and the Average Treatment Effect for the Overlap population (ATO) [3]. For variance estimation, robust variance estimators that do not account for weight estimation (hereafter, robust variance) are frequently used and serve as the default in many statistical software packages [1,4–6]. For continuous outcomes, it is known that the robust variance estimator is conservative for the ATE; however, recent studies report that this property does not always hold for other estimands such as the ATT or ATO [7,8].

In medical research, time-to-event outcomes are frequently analyzed. In the context of propensity score weighted Cox models (CoxPSW) [9], the robust variance has been shown to be conservative when estimating the ATE [10]. However, the properties of the robust variance estimator in CoxPSW for other estimands, such as the ATT and ATO, remain unclear. In this study, we investigate the properties of the robust variance estimator in CoxPSW for estimands other than the ATE through mathematical derivation, simulation studies, and data analysis.

# 2. Extension of the sandwich variance to weighting schemes other than the ATE

We extend the asymptotic variance estimator accounting for weight estimation proposed by Shu et al. [10] to a general class of weights beyond the ATE.

Suppose we observe a random sample of $n$ individuals from a target population. For each individual $i$, let $\boldsymbol{X_i}$ denote a vector of baseline characteristics, and let $A_i \in \{0,1\}$ denote an indicator whether the individual received the treatment or not. Each individual is followed until either an event of interest or another event, denoted as censoring, occurs, and the observed time is defined as

$$T_i = \min(T_i^*, C_i),$$

where $T_i^*$ represents the event time and $C_i$ the censoring time. We further define $\delta_i = I(T_i^* \leq C_i)$ as the event indicator. Throughout this work, we assume that, conditional on treatment assignment $A_i$, the censoring mechanism does not depend on the underlying event time $T_i^*$ nor the covariates $X_i$. The observed data for each individual are $(X_i A_i, T_i, \delta_i)$. We assume that the standard causal inference assumptions hold: consistency, positivity, and conditional exchangeability.

Let $T_a^*$ denote the counterfactual time to failure that would have been observed for an individual had the treatment level been set to $A_i = a$. We assume that $T_a^*$ follows the proportional hazards model given below:

$$\lambda_a(t) = \lambda_0(t) \exp(\theta a),$$

where $\lambda_a(t)$ denotes the hazard function for the counterfactual failure time. The parameter $\theta$ presents the marginal log hazard ratio comparing the hazard under treatment with that under control. To obtain an estimate of $\theta$, we adopt a PSW approach that reweights the observed sample. The estimator $\hat{\theta}$ is defined as the solution to the weighted score equation derived from the partial likelihood [11]:

$$\sum_{i=1}^{n} \widehat{w}_i \delta_i \left[ A_i - \frac{\sum_{\ell \in \mathcal{R}_i} \widehat{w}_\ell \exp(\theta A_\ell) A_\ell}{\sum_{\ell \in \mathcal{R}_i} \widehat{w}_\ell \exp(\theta A_\ell)} \right] = 0.$$

Here, $\mathcal{R}_i = \{\ell: T_\ell \geq T_i, \delta_i = 1\}$ denotes the risk set at the event time of individual $i$, and $\hat{w}_i$ represents the estimated weight for individual $i$.

Let the weight be defined as a general function of the estimated propensity score, where the propensity score is defined as $e_i = P(A_i = 1|X_i)$:

$$w_i = \frac{w(e_i)}{A_i e_i + (1 - A_i)(1 - e_i)}.$$

Here, $w(e_i)$ denotes an estimand-specific weight function evaluated at the propensity score [3]. For instance, $w(e_i) = 1$ for the ATE, $w(e_i) = e_i$ for the ATT, and $w(e_i) = e_i(1 - e_i)$ for the ATO.

To account for the estimation of the propensity score, we consider a sandwich variance estimator that incorporates uncertainty from the design stage [12]. Following Shu et al., let $\gamma$ denote the parameter vector in the propensity score model. Define $\boldsymbol{\beta} = (\theta, \boldsymbol{\gamma}^\top)^\top$. The joint estimating equation for $\boldsymbol{\beta}$ can be written as

$$\sum_{i=1}^n \Phi_i(\theta, \boldsymbol{\gamma}) = \begin{pmatrix} \sum_{i=1}^n \psi_i(\theta, \boldsymbol{\gamma}) \\ \sum_{i=1}^n \pi_i(\boldsymbol{\gamma}) \end{pmatrix} = \begin{pmatrix} \sum_{i=1}^n w_i \delta_i \left[ A_i - \frac{\sum_{\ell \in \mathcal{R}_i} w_\ell \exp(\theta A_\ell) A_\ell}{\sum_{\ell \in \mathcal{R}_i} w_\ell \exp(\theta A_\ell)} \right] \\ \sum_{i=1}^n \left[ A_i - \frac{1}{\{1 + \exp(-\boldsymbol{\gamma}^T X_i)\}} \right] X_i \end{pmatrix} = \mathbf{0}$$

where the propensity score is estimated using the score function $\pi_i(\boldsymbol{\gamma})$ from the logistic model with 1 included in the vector of covariates $X_i$. The asymptotic variance estimator (hereafter, corrected sandwich variance) of $\hat{\boldsymbol{\beta}} = (\hat{\theta}, \hat{\boldsymbol{\gamma}}^\top)^\top$ is given by the following expression. The derivation follows the same steps as in Appendix A of Shu et al., which applies the arguments of Lin and Wei (1989) and Binder (1992) [13,14].

The corrected sandwich variance estimator is given by

$$\widehat{var}_{CS}(\hat{\boldsymbol{\beta}}) = A(\hat{\boldsymbol{\beta}})^{-1} B(\hat{\boldsymbol{\beta}}) \{A(\hat{\boldsymbol{\beta}})^{-1}\}^\top,$$

Where $A(\widehat{\boldsymbol{\beta}}) = -\sum_{i=1}^{n} \Phi'_i(\widehat{\boldsymbol{\beta}})$ and $B(\widehat{\boldsymbol{\beta}}) = \sum_{i=1}^{n} \Omega_i(\widehat{\boldsymbol{\beta}})\Omega_i(\widehat{\boldsymbol{\beta}})^{\top}$. Here, the prime in the definition of $A(\widehat{\boldsymbol{\beta}})$ denotes the first derivative. $\Omega_i(\widehat{\boldsymbol{\beta}})$ is defined as

$$\Omega_i(\widehat{\boldsymbol{\beta}}) = \begin{pmatrix} \eta_i(\widehat{\theta}, \widehat{\boldsymbol{\gamma}}) \\ \pi_i(\widehat{\boldsymbol{\gamma}}) \end{pmatrix},$$

where

$$\eta_i(\widehat{\theta}, \widehat{\boldsymbol{\gamma}}) = \widehat{w}_i \, \delta_i \left\{ A_i - \frac{S_1(i)}{S_0(i)} \right\} - \widehat{w}_i A_i \exp(A_i\widehat{\theta}) \sum_{j=1}^{n} \delta_j \, \widehat{w}_j \frac{I(T_j \le T_i)}{S_0(j)}$$

$$+ \widehat{w}_i \exp(A_i\widehat{\theta}) \sum_{j=1}^{n} \delta_j \, \widehat{w}_j I(T_j \le T_i) \frac{S_1(j)}{S_0(j)^2},$$

$$S_0(i) = \sum_{l \in \mathcal{R}_i} \widehat{w}_l \exp(A_l \widehat{\theta}),$$

$$S_1(i) = \sum_{l \in \mathcal{R}_i} \widehat{w}_l \exp(A_l \widehat{\theta}) A_l$$

Corrected sandwich variance is essentially identical to that presented by Zhao et al. for the case of a continuous treatment variable [15].

## 3. Conservativeness under non-ATE weighting

As noted above, when using ATE weights $w(e_i) \equiv 1$, robust variance is conservative. In the Supplementary File, we show that for general weight functions other than those corresponding to the ATE, robust variance is not necessarily asymptotically conservative. This finding is consistent with previous results for continuous outcomes [7,8].

## 4. Simulation study

Following the simulation framework for continuous outcomes [8], we generated 1,000 independent datasets, each consisting of n = 1,000 or 10,000 independent and identically distributed observations. For each dataset, we simulated three covariates (*L*1, *L*2, *L*3), a treatment indicator (*A*), event status ($\delta$) and a recorded time ($T$).

$$L1_i \sim Ber(0.5), L2_i \sim N(0,1), L3_i \sim N(0,1),$$

$$A_i \sim Ber\left(\frac{1}{1+\exp(2+0.5L1_i-\alpha_1 L2_i - \alpha_2 L3_i)}\right),$$

$$T_{0i} = Exp(0.01),$$

$$\eta_i = \log(0.8)A_i + \log(0.4)L1_i + \log(5)A_i L1_i + \log(\beta_1)L2_i + \log(\beta_2)L3_i,$$

$$T_i^* = T_{0i}\exp(-\eta_i),$$

$$C_i = Exp(0.0001),$$

$$\delta_i = I(T_i^* \leq C_i),$$

$$T_i = \min(T_i^*, C_i)$$

$\alpha_1, \alpha_2, \beta_1, \beta_2$ denote parameters that differ across simulation settings. The two scenarios are presented in Table 1. In the first scenario, confounders were more strongly associated with the outcome than treatment, whereas in the second scenario, confounders were more strongly associated with the treatment than outcome. The censoring rate was approximately 2%.

[Insert Table 1 here]

For each scenario presented in Table 1, CoxPSW were applied to estimate ATE, ATT, and ATO. For variance estimation, we compared a robust variance estimator with a corrected sandwich variance estimator in terms of coverage probability and the width of the 95% confidence interval (CI). The true hazard ratio was approximated by the estimated hazard ratio obtained from 50,000,000 samples under no censoring for each estimand.

The results are presented in Table 2 and 3. In the first scenario, the robust variance estimator was conservative regardless of the weighting scheme used. In contrast, in the second scenario, the coverage probability of the robust variance estimator fell well below 95% when ATT (0.905) and ATO (0.919) weights were used. The corrected sandwich variance achieved approximately 95% coverage. Similar results were observed for CI width. In the first scenario, the robust variance estimator yielded wider 95% CIs on average across all weighting schemes. In the second scenario, the 95% CIs based on the robust variance estimator were narrower for ATT and ATO, whereas they were slightly wider for the ATE.

[Insert Table2, Table3 here]

**5. Real data analysis**

In this section, we provide an illustrative example using real-world data on muscle-invasive bladder cancer [16]. Our analysis focused on estimating the causal effects of adjuvant therapy on cancer-specific mortality. The study sample consisted of 322 patients, including 74 in the adjuvant chemotherapy group (treatment group) and 248 in the radical cystectomy alone group (control group). Cancer-specific death occurred in 23 patients (31.1%) in the treatment group and in 45 patients (18.2%) in the control group. In this study, we conducted analyses for the overall population as well as subgroup analyses [16]. For each subgroup, analyses were performed using ATE, ATT, and ATO weighting. Variances were estimated using the robust variance estimator and the corrected sandwich variance estimator.

The results are presented in Table 4. For the overall population analysis, the robust variance estimator yielded more conservative results across all three weighting schemes. In contrast, in the subgroup analyses, there were several subgroups in which the robust variance estimator produced smaller variance estimates for ATT or ATO weighting compared with the corrected sandwich variance.

[Insert Table 4 here]

**6. Discussion**

In this study, we showed that the robust variance estimator that ignores the estimation of weights in CoxPSW is not necessarily conservative for general weighting schemes other than the ATE. This result is consistent with previous findings for continuous outcomes [7,8].

The simulation results demonstrated that the robust variance estimator may fail to be conservative when ATT or ATO weights are used. A general pattern observed in the simulations was that when covariates were weakly associated with the exposure but strongly associated with the outcome, the robust variance estimator tended to be conservative regardless of the weighting scheme. Similar phenomena related to prognostic variables have been reported for continuous outcomes [17]. Further research is needed to clarify the theoretical properties of the asymptotic variance of CoxPSW.

In the real data analysis, we also compared the robust variance estimator with the corrected sandwich variance estimator. Because the theory of the robust variance estimator is based on asymptotic variance, its properties are not necessarily guaranteed in finite samples. In the present data analysis, when ATE weights were used, the robust variance estimator produced larger estimates across all subgroups, as expected. In contrast, when ATT or ATO weights were used, the robust variance estimator produced smaller estimates in some subgroups.

In conclusion, similar to the cases of continuous outcomes, the robust variance estimator in CoxPSW is not guaranteed to be conservative when weighting schemes other than the ATE are used. Therefore, when applying CoxPSW, it is preferable to use variance estimators that account for the estimation of the weights.

**Acknowledgement**

The authors have nothing to report.

**Table1** Simulation scenario

| Scenario | $\alpha_1$ | $\alpha_2$ | $\beta_1$ | $\beta_2$ |
|---|---|---|---|---|
| Scenario 1 | -0.1 | -0.1 | 0.4 | 0.4 |
| Scenario 2 | 0.5 | 0.5 | 0.95 | 0.95 |

**Table2** Simulation results for scenario 1

| Weight | N | True log HR | Bias of log HR | Mean confidence interval width | | Coverage probability | |
|---|---|---|---|---|---|---|---|
| | | | | Robust variance | Corrected variance | Robust variance | Corrected variance |
| ATE | 1000 | 0.3145 | 0.00248 | 0.414 | 0.296 | 0.991 | 0.942 |
| ATE | 10000 | 0.3145 | 0.00218 | 0.130 | 0.092 | 0.992 | 0.946 |
| ATT | 1000 | 0.2219 | -0.00038 | 0.430 | 0.355 | 0.986 | 0.949 |
| ATT | 10000 | 0.2219 | -0.00049 | 0.137 | 0.113 | 0.978 | 0.944 |
| ATO | 1000 | 0.2326 | -0.00028 | 0.425 | 0.344 | 0.976 | 0.936 |
| ATO | 10000 | 0.2326 | 0.00181 | 0.135 | 0.109 | 0.986 | 0.948 |

Abbreviations: Corrected variance, corrected sandwich variance; HR, hazard ratio; log, natural logarithm; N, sample size

**Table3** Simulation results for scenario 2

| Weight | N | True log HR | Bias of log HR | Mean confidence interval width | | Coverage probability | |
|---|---|---|---|---|---|---|---|
| | | | | Robust variance | Corrected variance | Robust variance | Corrected variance |
| ATE | 1000 | 0.5450 | -0.00190 | 0.421 | 0.406 | 0.955 | 0.946 |
| ATE | 10000 | 0.5450 | -0.00003 | 0.135 | 0.132 | 0.949 | 0.946 |
| ATT | 1000 | 0.4006 | 0.00180 | 0.397 | 0.455 | 0.912 | 0.951 |
| ATT | 10000 | 0.4006 | -0.00133 | 0.126 | 0.144 | 0.905 | 0.953 |
| ATO | 1000 | 0.4247 | -0.00165 | 0.389 | 0.429 | 0.930 | 0.952 |
| ATO | 10000 | 0.4247 | -0.00126 | 0.123 | 0.136 | 0.919 | 0.952 |

Abbreviations: Corrected variance, corrected sandwich variance; HR, hazard ratio; log, natural logarithm; N, sample size

**Table 4** Results of the data analysis

| Weight | Subgroup | log HR | Robust SE | Corrected sandwich SE | Robust SE /Corrected sandwich SE |
|---|---|---|---|---|---|
| ATE | Overall population | -0.148 | 0.287 | 0.237 | 1.211 |
| ATE | Age (years) 70>= | -0.661 | 0.440 | 0.366 | 1.202 |
| ATE | Age (years) 70< | -0.161 | 0.419 | 0.380 | 1.103 |
| ATE | Male | -0.486 | 0.329 | 0.260 | 1.265 |
| ATE | Female | 0.162 | 0.736 | 0.703 | 1.047 |
| ATE | NAC present | 0.260 | 0.618 | 0.538 | 1.149 |
| ATE | NAC absent | -0.505 | 0.330 | 0.261 | 1.264 |
| ATE | Pathological T stage 3>= | -0.421 | 0.287 | 0.259 | 1.108 |
| ATE | Pathological T stage 2<= | -1.038 | 0.687 | 0.618 | 1.112 |
| ATE | Lymph nodes metastasis present | -1.569 | 0.404 | 0.393 | 1.028 |
| ATE | Lymph nodes metastasis absent | -0.121 | 0.379 | 0.302 | 1.255 |
| ATE | Venous invasion present | -0.701 | 0.336 | 0.288 | 1.167 |
| ATE | Venous invasion absent | -0.399 | 0.541 | 0.495 | 1.093 |
| ATE | Lymphatic duct invasion present | -0.760 | 0.318 | 0.272 | 1.169 |
| ATE | Lymphatic duct invasion absent | -0.010 | 0.553 | 0.503 | 1.099 |
| ATT | Overall population | -0.485 | 0.257 | 0.237 | 1.084 |
| ATT | Age (years) 70>= | -0.731 | 0.334 | 0.321 | 1.040 |
| ATT | Age (years) 70< | -0.416 | 0.360 | 0.339 | 1.062 |
| ATT | Male | -0.543 | 0.272 | 0.250 | 1.088 |
| ATT | Female | -0.101 | 0.606 | 0.611 | 0.992 |
| ATT | NAC present | -0.228 | 0.617 | 0.588 | 1.049 |
| ATT | NAC absent | -0.543 | 0.275 | 0.250 | 1.100 |
| ATT | Pathological T stage 3>= | -0.647 | 0.279 | 0.267 | 1.045 |
| ATT | Pathological T stage 2<= | -0.361 | 0.670 | 0.668 | 1.003 |
| ATT | Lymph nodes metastasis present | -1.733 | 0.426 | 0.421 | 1.012 |
| ATT | Lymph nodes metastasis absent | 0.149 | 0.343 | 0.326 | 1.052 |
| ATT | Venous invasion present | -0.765 | 0.283 | 0.278 | 1.018 |
| ATT | Venous invasion absent | 0.126 | 0.492 | 0.510 | 0.965 |
| ATT | Lymphatic duct invasion present | -0.872 | 0.287 | 0.281 | 1.021 |
| ATT | Lymphatic duct invasion absent | 0.753 | 0.511 | 0.531 | 0.962 |
| ATO | Overall population | -0.322 | 0.260 | 0.233 | 1.116 |
| ATO | Age (years) 70>= | -0.619 | 0.339 | 0.309 | 1.097 |
| ATO | Age (years) 70< | -0.145 | 0.377 | 0.355 | 1.062 |
| ATO | Male | -0.38 | 0.277 | 0.244 | 1.135 |
| ATO | Female | 0.246 | 0.613 | 0.617 | 0.994 |
| ATO | NAC present | 0.094 | 0.598 | 0.527 | 1.135 |
| ATO | NAC absent | -0.372 | 0.281 | 0.249 | 1.129 |
| ATO | Pathological T stage 3>= | -0.428 | 0.282 | 0.256 | 1.102 |
| ATO | Pathological T stage 2<= | -0.335 | 0.650 | 0.631 | 1.030 |
| ATO | Lymph nodes metastasis present | -1.621 | 0.421 | 0.410 | 1.027 |
| ATO | Lymph nodes metastasis absent | 0.185 | 0.335 | 0.308 | 1.088 |
| ATO | Venous invasion present | -0.605 | 0.299 | 0.271 | 1.103 |
| ATO | Venous invasion absent | 0.127 | 0.482 | 0.484 | 0.996 |
| ATO | Lymphatic duct invasion present | -0.705 | 0.296 | 0.265 | 1.117 |
| ATO | Lymphatic duct invasion absent | 0.660 | 0.503 | 0.501 | 1.004 |

Abbreviations: HR, hazard ratio; log, natural logarithm; N, sample size; NAC, neoadjuvant chemotherapy; SE, standard error

**Supplemental File**

**Comparison of Asymptotic Variances under General Weights**

In Shu et al.[1], it is shown that for CoxPSW using ATE weights, the robust variance is asymptotically larger than the corrected sandwich variance, meaning that the robust variance is conservative. Here, we show that for other types of weights, the robust variance is not necessarily conservative in general. The notation used in this study follows Appendix A–C of Shu et al. Furthermore, the relationship between robust variance and corrected sandwich variance used in the proof is based on the result in Appendix B of Shu et al.

We denote the robust variance estimator by $\widehat{var}_R(\hat{\theta})$ and the corrected sandwich variance estimator by $\widehat{var}_{CS}(\hat{\beta})$. The estimator $\widehat{var}_{CS}(\hat{\beta})$ is defined as described in the main text. The robust variance estimator $\widehat{var}_R(\hat{\theta})$ is given by the following expression.

$$\widehat{var}_R(\hat{\theta}) = A(\hat{\theta})^{-1} B(\hat{\theta}) \left\{ A(\hat{\theta})^{-1} \right\}^\mathrm{T}$$

where

$$A(\hat{\theta}) = -\sum_{i=1}^n \psi'_i(\hat{\theta}),$$

$$B(\hat{\theta}) = \sum_{i=1}^n \eta_i(\hat{\theta}) \eta_i(\hat{\theta})^\mathrm{T},$$

$$\eta_i(\hat{\theta}) = \hat{w}_i \, \delta_i \left\{ A_i - \frac{S_1(i)}{S_0(i)} \right\} - \hat{w}_i A_i \exp(A_i \hat{\theta}) \sum_{j=1}^n \delta_j \, \hat{w}_j \frac{I(T_j \leq T_i)}{S_0(j)}$$

$$+ \hat{w}_i \exp(A_i \hat{\theta}) \sum_{j=1}^n \delta_j \, \hat{w}_j I(T_j \leq T_i) \frac{S_1(j)}{S_0(j)^2},$$

$$S_0(i) = \sum_{l \in \mathcal{R}_i} \hat{w}_l \exp(A_l \hat{\theta}),$$

$$S_1(i) = \sum_{l \in \mathcal{R}_i} \hat{w}_l \exp(A_l \hat{\theta}) A_l.$$

Here, the prime in the definition of $A(\hat{\theta})$ denotes the first derivative.

The relationship between correct sandwich variance and robust variance can be expressed as:

$$n\widehat{var}_{CS}(\hat{\theta}) = n\widehat{var}_R(\hat{\theta}) + \frac{n^2}{(A_{11})^2}\left[\hat{d}^\top \frac{1}{n}\sum_{i=1}^n\{(A_i - \hat{e}_i^2)X_iX_i^\top\}\hat{d} + 2\hat{d}^\top \frac{1}{n}\sum_{i=1}^n \hat{w}_i L_{0i}(A_i - \hat{e}_i)X_i\right]$$

where $A(\hat{\beta}) = -\sum_{i=1}^n \Phi'_i(\hat{\beta}) = -\sum_{i=1}^n \begin{bmatrix} \frac{\partial \psi_i(\hat{\beta})}{\partial \hat{\theta}} & \frac{\partial \psi_i(\hat{\beta})}{\partial \hat{\gamma}^\top} \\ \frac{\partial \pi_i(\hat{\gamma})}{\partial \hat{\theta}} & \frac{\partial \pi_i(\hat{\gamma})}{\partial \hat{\gamma}^\top} \end{bmatrix} = \begin{bmatrix} A_{11} & A_{12} \\ 0 & A_{22} \end{bmatrix}$,

where $\widehat{var}_{CS}(\hat{\theta})$ is given by the element in the first row and first column of the matrix $\widehat{var}_{CS}(\hat{\beta})$,

$\hat{d} = \hat{U}^{-1}\{\frac{1}{n}\sum_{i=1}^n k_i \hat{L}_{0i} X_i\}$, $k_i = A_i(1-\hat{e}_i)\left(w(\hat{e}_i)' - \frac{w(\hat{e}_i)}{\hat{e}_i}\right) + (1-A_i)\hat{e}_i\left(w(\hat{e}_i)' + \frac{w(\hat{e}_i)}{1-\hat{e}_i}\right)$,

$\hat{U} = \frac{1}{n}\sum_{i=1}^n \hat{e}_i(1-\hat{e}_i)X_i X_i^\top$, and $L_{0i} = \delta_i\left\{A_i - \frac{S_1(i)}{S_2(i)}\right\} - \exp(\hat{\theta}A_i)[\sum_{j=1}^n \frac{\hat{w}_j \delta_j I(T_j \leq T_i)}{S_0(j)}\{A_i - \frac{S_1(i)}{S_2(i)}\}$

Here, which asymptotic variance is larger is determined by the sign of the final term. Rearranging the part that determines the sign of the final term in the same way as in Shu et al., we obtain the following expression:

$$\hat{d}^\top \frac{1}{n}\sum_{i=1}^n\{(A_i - \hat{e}_i^2)X_iX_i^\top\}\hat{d} + 2\hat{d}^\top \frac{1}{n}\sum_{i=1}^n \hat{w}_i L_{0i}(A_i - \hat{e}_i)X_i$$

$$= \left\{\frac{1}{n}\sum_{i=1}^n k_i\hat{L}_{0i}X_i\right\}^\top \hat{U}^{-1} \frac{1}{n}\sum_{i=1}^n\{(A_i - \hat{e}_i^2)X_iX_i^\top\}\hat{U}^{-1}\left\{\frac{1}{n}\sum_{i=1}^n k_i\hat{L}_{0i}X_i\right\}$$

$$+ 2\left\{\frac{1}{n}\sum_{i=1}^n k_i\hat{L}_{0i}X_i\right\}^\top \hat{U}^{-1} \frac{1}{n}\sum_{i=1}^n \hat{w}_i \hat{L}_{0i}(A_i - \hat{e}_i)X_i$$

$$\approx \left\{\frac{1}{n}\sum_{i=1}^n k_i\hat{L}_{0i}X_i\right\}^\top \hat{U}^{-1}\left\{\frac{1}{n}\sum_{i=1}^n k_i\hat{L}_{0i}X_i\right\} + 2\left\{\frac{1}{n}\sum_{i=1}^n k_i\hat{L}_{0i}X_i\right\}^\top \hat{U}^{-1}\frac{1}{n}\sum_{i=1}^n \hat{w}_i\hat{L}_{0i}(A_i - \hat{e}_i)X_i \quad (1)$$

where $A_n \approx B_n$ means that $A_n - B_n = o_p(1)$ for sequences of random variables $A_n$ and $B_n$. We used $U_n \approx \frac{1}{n}\sum_{i=1}^n\{(A_i - \hat{e}_i^2)X_iX_i^\top\}$.

In the ATE case, the relationship $\hat{w}_i(A_i - \hat{e}_i) = -k_i$ holds. In the case of the ATE, this expression simplifies to the following form. Equation (1) can be rearranged as follows.

$$-\left\{\frac{1}{n}\sum_{i=1}^{n} k_i \hat{L}_{0i} X_i\right\}^{\top} \hat{U}^{-1} \left\{\frac{1}{n}\sum_{i=1}^{n} k_i \hat{L}_{0i} X_i\right\} \leq 0$$

Therefore, the expression becomes a quadratic form, which is always non-positive. However, for weights other than ATE, the sign of the above expression is not determined in general.

For example, the expressions for the ATT and ATO are given as follows. In the case of the ATT, $w(\hat{e}_i) = \hat{e}_i$. Then, $k_i = A_i(1-\hat{e}_i)\left(w(\hat{e}_i)' - \frac{w(\hat{e}_i)}{\hat{e}_i}\right) + (1-A_i)\hat{e}_i\left(w(\hat{e}_i)' + \frac{w(\hat{e}_i)}{1-\hat{e}_i}\right) = (1-A_i)\frac{\hat{e}_i}{(1-\hat{e}_i)}$, and $\hat{w}_i(A_i - \hat{e}_i) = A(1-\hat{e}_i) - (1-A)(\frac{\hat{e}^2_i}{1-\hat{e}_i})$. Thus, the sign is not uniquely determined.

In the case of the ATO, $w(\hat{e}_i) = \hat{e}_i(1-\hat{e}_i)$. Then, $k_i = -A_i\hat{e}_i(1-\hat{e}_i) + (1-A_i)\hat{e}_i(1-\hat{e}_i)$, and $\hat{w}_i(A_i - \hat{e}_i) = A(1-\hat{e}_i) - (1-A)(\frac{\hat{e}^2_i}{1-\hat{e}_i})$. Thus, in this case as well, the sign is not uniquely determined.

Consequently, the relative magnitude of the asymptotic variances depends on the specific setting.